\documentclass[longauth]{aa} 
\usepackage{graphicx}
\usepackage{lscape}
\usepackage{booktabs}
\usepackage{longtable}
\usepackage{multicol}
\usepackage{multirow}
\usepackage{txfonts}
\usepackage{natbib}
\usepackage{subfigure}
\usepackage{amsmath}
\usepackage{color}
\usepackage{url}
\usepackage[pdfpagemode=UseThumbs,colorlinks=true,bookmarks=true]{hyperref}
\hypersetup{citecolor={blue},linkcolor={blue},urlcolor={blue}}
\bibpunct{(}{)}{;}{a}{}{,}

\begin{document} 

    \title{Central star formation and metallicity in CALIFA interacting galaxies}
\titlerunning{Central sSFR and Z in CALIFA interacting galaxies}
                        \author{
                        J.K.~Barrera-Ballesteros \inst{\ref{iac},\ref{ull}} \fnmsep\thanks{\email{jkbb@iac.es}}
                        \and
                        S. F.~S\'anchez \inst{\ref{unam}}
                        \and
                        B. Garc\'ia-Lorenzo \inst{\ref{iac},\ref{ull}}
                        \and
                        J. Falc\'on-Barroso \inst{\ref{iac},\ref{ull}}
                        \and
                        D. Mast \inst{\ref{ICRA}}
                        \and
                        R. Garc\'ia-Benito \inst{\ref{iaa}}
                        \and
                        B. Husemann \inst{\ref{eso}}
                        \and
                        G. van de Ven \inst{\ref{mpia}}
                        \and
                        J. Iglesias-P\'aramo \inst{\ref{iaa},\ref{arido}}
                        \and
                        F.F. Rosales-Ortega \inst{\ref{INAOE}}
                        \and
                        M.A. P\'erez-Torres \inst{\ref{iaa},\ref{teruel}}
                        \and
                        I.~M\'{a}rquez \inst{\ref{iaa}}
                        \and
                        C.~Kehrig \inst{\ref{iaa}}
                   \and 
         R.\,A.~Marino\inst{\ref{ucm}}
                        \and
                        J. M. Vilchez \inst{\ref{iaa}}
                        \and
                        L. Galbany \inst{\ref{mile-chile},\ref{UdC}}
                        \and
                        \'A.R. L\'opez-S\'anchez \inst{\ref{aao},\ref{macquireUni}}
                        \and
                        C.\,J.\,~Walcher\inst{\ref{aip}}
                        \and
                        the~CALIFA~collaboration}

\institute{
\label{iac}Instituto de Astrof\'\i sica de Canarias (IAC), E-38205 La Laguna, Tenerife, Spain
\and
\label{ull}Depto. Astrof\'\i sica, Universidad de La Laguna (ULL), E-38206 La Laguna, Tenerife, Spain
\and
\label{unam}Instituto de Astronom\'\i a,Universidad Nacional Auton\'oma de Mexico, A.P. 70-264, 04510, M\'exico,D.F.
\and
\label{ICRA} Instituto de Cosmologia, Relatividade e Astrof\'{i}sica – ICRA, Centro Brasileiro de Pesquisas F\'{i}sicas, Rua Dr.Xavier Sigaud 150, CEP 22290-180, Rio de Janeiro, RJ, Brazil
\and
\label{iaa}Instituto de Astrof\'{\i}sica de Andaluc\'{\i}a (CSIC), Glorieta de la Astronom\'\i a s/n, Aptdo. 3004, E18080-Granada, Spain
\and
\label{eso} European Southern Observatory (ESO), Karl-Schwarzschild-Str. 2, D-85748 Garching b. Muenchen, Germany
\and
\label{mpia}Max-Planck-Institut f\"ur Astronomie, Heidelberg, Germany.
\and
\label{arido}Estaci\'{o}n Experimental de Zonas Aridas (CSIC), Ctra. de Sacramento s/n, La Ca\~{n}ada, Almer\'{\i}a, Spain
\and
\label{INAOE}Instituto Nacional de Astrof\'isica, \'Optica y Electr\'onica, Luis E. Erro 1, 72840 Tonantzintla, Puebla, Mexico
\and
\label{teruel}Centro de Estudios de la F\'isica del Cosmos de Arag\'on, E-44001 Teruel, Spain
\and
\label{ucm}CEI Campus Moncloa, UCM-UPM, Departamento de Astrof\'{i}sica y CC$.$ de la Atm\'{o}sfera, Facultad de CC$.$ F\'{i}sicas, Universidad Complutense de Madrid, Avda.\,Complutense s/n, 28040 Madrid, Spain.
\and
\label{mile-chile}Millennium Institute of Astrophysics, Universidad de Chile, Santiago, Chile
\and
\label{UdC}Departamento de Astronom\'ia, Universidad de Chile, Casilla 36-D, Santiago, Chile
\and
\label{aao}Australian Astronomical Observatory, PO Box 915, North Ryde, NSW 1670, Australia
\and
\label{macquireUni}Department of Physics and Astronomy, Macquarie University , NSW 2109, Australia
\and
\label{aip} Leibniz-Institut f\"ur Astrophysik Potsdam (AIP), An der Sternwarte 16, D-14482 Potsdam, Germany
}          
 \abstract
   {We use optical integral-field spectroscopic (IFS) data from 103 nearby galaxies at different stages of the merging event, from close pairs to merger remnants provided by the CALIFA survey, to study the impact of the interaction in the specific star formation and oxygen abundance on different galactic scales. To disentangle the effect of the interaction and merger from internal processes, we compared our results with a control sample  of 80 non-interacting galaxies. We confirm the moderate enhancement ($\times$ 2-3 times) of specific star formation for interacting galaxies in central regions as reported by previous studies; however, the specific star formation is comparable when observed in extended regions. We find that control and interacting star-forming galaxies have similar oxygen abundances in their central regions, when normalized to their stellar masses. Oxygen abundances of these interacting galaxies seem to decrease compared to the control objects at the large aperture sizes measured in effective radius. Although the enhancement in central star formation and lower metallicities for interacting galaxies have been attributed to tidally induced inflows, our results suggest that other processes such as stellar feedback can contribute to the metal enrichment in interacting galaxies.}

\maketitle

\section{Introduction}

Interactions and mergers are identified as key mechanisms in increasing the star formation rate (SFR) in galaxies \citep[e.g.,][]{1996ARA&A..34..749S, 1999Ap&SS.266..137B}. In particular, luminous infrared galaxies (LIRGs), which are associated almost exclusively to merger events, present strong episodes of star formation \citep[e.g.,][]{2013MNRAS.430.3128E}.
The scenario proposed by these studies and hydrodynamical numerical simulations \citep[e.g.,][]{1996ApJ...464..641M, 1996ApJ...471..115B} suggests that this enhancement is the result of tidally induced inflows of gas that favors the increment of star formation activity in the central region of interacting and merging galaxies.
On the other hand, large samples of interacting and merging galaxies in the Sloan Digital Sky Survey \citep[SDSS, e.g.,][]{2008AJ....135.1877E,2013MNRAS.435.3627E,2012MNRAS.426..549S,  2013MNRAS.433L..59P} indicate that although central SFR is enhanced in these objects compared with non-interacting galaxies of similar stellar mass, on average this increment is rather moderate. This indicates that the triggering of starburst episodes depends on more parameters than the mere fact of galaxies being in interaction \citep[orbital configuration and intrinsic properties of the progenitors; e.g., ][]{2006MNRAS.373.1013C,2008A&A...492...31D}.

It is also worthwhile noting that statistical spectroscopic studies like the one above are performed by observing a fixed projected portion of the galaxies \citep[e.g., SDSS or GAMA fiber sizes;][]{2014MNRAS.445.1157C,2014MNRAS.444.3986R}. This yields measurements on different galactic scales, making it difficult to assess the extension of star formation. In this regard, IFS observations of interacting galaxies are crucial for understanding whether the enhancement in the SFR is a global or a localized process. An example of the usefulness of such observations has been demonstrated by \cite{2014A&A...567A.132W}. From a detailed study of the Mice galaxies, they found that there is a moderate increment in the central SFR and no net enhancement in the global SFR. This study considers a single interacting system. Statistical studies of spatially resolved star formation in interacting galaxies are therefore required. Even more, to make a fair comparison, a homogeneously observed sample of non-interacting galaxies is also required. Recently, numerical simulations are starting to explore the extend of star formation induced by the interaction. \cite{2015MNRAS.448.1107M} find enhancement in the star formation on the central kpc scales and moderately suppressed activity at larger galacto-centric radii.

In the above picture, these metal-poor gas inflows decrease the central metallicity in comparison to non-interacting galaxies \citep[e.g.,][]{2006AJ....131.2004K}. As a consequence, merging galaxies are thought to contribute to the scatter in the mass-metallicity relation (hereafter M-Z relation) of star forming galaxies. However, recent numerical simulations that included feedback processes have suggested that nuclear metallicity depends on more factors than the dynamics of the gas, such as the chemical enrichment due to the ongoing star formation, the stellar (and possible AGN) feedback, and returned material by evolved stars into the interstellar medium (ISM) of the entire galaxy \citep[][]{2012ApJ...746..108T}. The result of the interplay between these different processes in these simulations could lead, in some cases, to a depression in the central metallicity or even an enhancement with respect to isolated galaxies. Spatially resolved observational studies are then required to shade some light on the chemical evolution of interacting or merging galaxies. 

Recently, IFS studies have been carried aimed at the understanding of properties of the ionized gas in individual \ion{H}{ii} regions for a large sample of galaxies. The radial gradient of oxygen abundance in galactic disks has been characterized, with an observed flattening in this gradient caused by interactions and mergers \citep{2014A&A...563A..49S}. They also confirmed the global and local nature of the M-Z relation \citep{2013A&A...554A..58S}. These studies have been possible thank to the integral-field spectroscopic (IFS) data provided by the CALIFA survey \citep{2012A&A...538A...8S}. CALIFA is an ongoing IFS survey aimed at obtaining spatially resolved physical properties, covering the full optical spatial extent (up to $\sim$ 2.5 times the effective radius), of 600 galaxies in the nearby universe (0.0005 $<$ z $<$ 0.03) of any morphological type. It is complete in a wide range of stellar masses \citep[ from $\sim$ 10$^{9.5}$ to $\sim$10$^{11}$M$_{\odot}$, ][]{2014A&A...569A...1W}. This dataset allows us to study for the first time spatially resolved properties of the ionized gas in a significant sample of interacting and merging galaxies to unveil the effect of interactions on different spatial scales and at different stages of the interaction. It also allows us to compare these properties with a sample of homogeneous observations of non-interacting galaxies.  

In this article we are aiming to investigate the impact of interactions in the star formation and oxygen abundance on different galactic scales. To accomplish this, we study the emission-line flux maps extracted from the data cubes of interacting and post-merger galaxies. From these maps we derived integrated properties, such as the H$\alpha$ equivalent width [EW(H$\alpha$) hereafter] and flux ratios essential for deriving oxygen abundances. This paper is organized as follows. In Section\,\ref{sec:data} we present the control and interacting samples, their ionized gas maps, and the aperture measurements. The comparison of these two samples on different galactic scales is presented in Section\,\ref{sec:results}. Finally,  we summarize our results in Section\,\ref{sec:conclusions}.  

\section{Sample and data}
\label{sec:data}
       
\subsection{Interacting and control samples}
\label{sec:Samples}
The galaxies that we use in this study are included in the CALIFA objects observed until May 2014. 
The selection of the interacting or merging sample is explained in a forthcoming article (Barrera-Ballesteros~et~al.,~submitted). Briefly, we first selected those objects with companions
from the observed CALIFA sample that are within a projected distance of 160 kpc and systemic velocities lower than $\Delta v$\,<\,600\,km\,s$^{-1}$. We noted that a large number of companion objects are not included in the CALIFA mother sample. To enhance the number of complete pairs in the survey, we conducted a complementary project aimed at observing companions not included in the original mother sample (P.I.s J.K. Barrera-Ballesteros and G. van de Ven). These observations and their reduction were carried out using the same strategy and pipeline as from the CALIFA survey. From the remaining sample of galaxies without close companions, we visually selected the objects with morphological evidence of merger event as post-merger systems. Such objects are morphologically disturbed, highly irregular, with evident tidal features or is there any signature of a recent merging event \citep{2013MNRAS.435.3627E}. The sample of interacting and merging consists of 103 objects. We compared our sample of interacting/merging galaxies with a control sample of 80 non-interacting galaxies to match the interacting sample in stellar mass and color. This sample is presented in \cite{2014A&A...568A..70B}, along with a comparison of their stellar and ionized gas kinematics.

\begin{figure*}[!htb]
 \includegraphics[width=\linewidth]{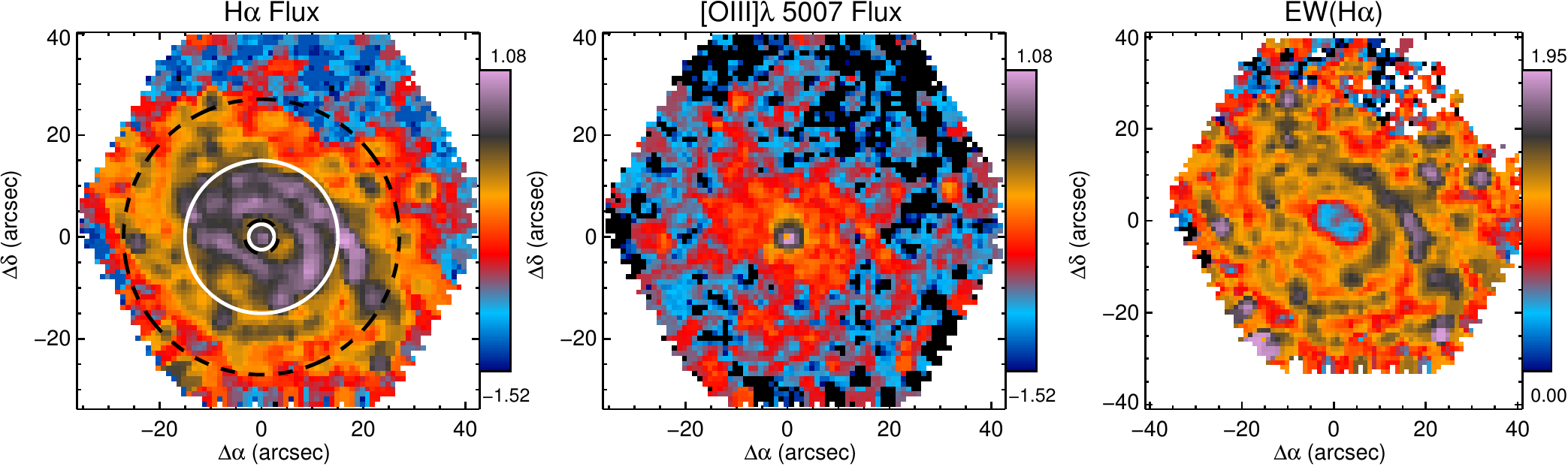}    
 \includegraphics[width=\linewidth]{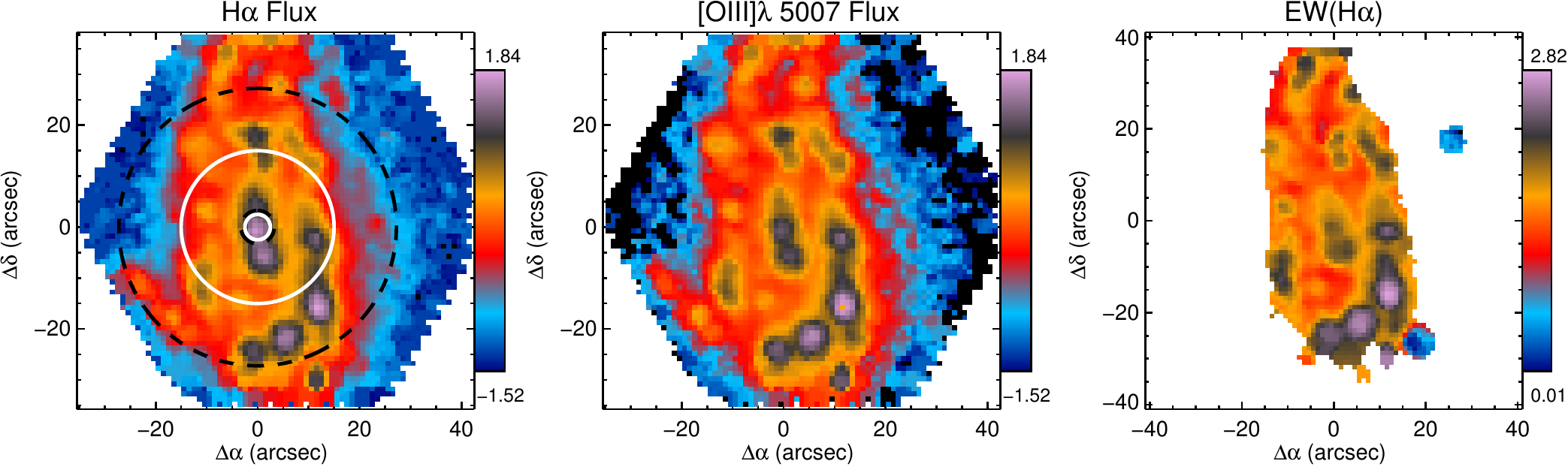}   
\caption{\label{map} Emission-lines properties of a galaxy in the control sample (top panels, NGC~0214) and in the interacting sample (bottom panels, UGC~312). Left panels show the H$\alpha$ flux. White-solid and black-dashed circles represent the central and extended apertures in arcsecs and effective radius units, respectively. Middle panel shows the [\ion{O}{iii}]$\lambda$ 5007, emission line flux. The units in both flux maps are (log10) 10$^{-16}$ erg s$^{-1}$ cm$^{-2}$ arcsec$^{-2}$. Right panels show the logarithm of the absolute values of EW(H$\alpha$) in units of \AA.}
\end{figure*}
\subsection{CALIFA data cubes and emission-line extraction}
\label{sec:Data}

Galaxies presented in this study were observed at the Calar Alto observatory using the PPAK instrument \citep{2005PASP..117..620R}. This instrument consists of 331 fibers of 2\farcs7 diameter each, concentrated in a single hegaxon bundle covering a field of view (FoV) of $74\arcsec\times64\arcsec$ with a filling factor of 60\%.  A three-point dithering technique is used to reach a filling factor of 100\% across the entire
FoV \citep[see details in][]{2013A&A...549A..87H}. For this study we used the data cubes with an intermediate nominal spectral resolution (V500 setup) of $\lambda$/$\Delta\lambda$\,$\sim$\,850 at $\sim$5000\AA. The wavelength range of the V500 setup covers several bright emission lines (3745$-$7300\AA): H$\alpha$, H$\beta$, H$\gamma$, [\ion{O}{ii}]~$\lambda\lambda$~3728,~3726,  [\ion{O}{iii}]~$\lambda\lambda$~4958,~5007, [\ion{N}{ii}]~$\lambda\lambda$~5755,~6584, [\ion{S}{ii}]~$\lambda\lambda$~6716,~6730, among others. The final data cube has an homogenized spectral resolution (FWHM) over the entire wavelength range of 6.0\AA,\  and the wavelength sampling per spaxel is 2.0\AA. The total exposure time per pointing is fixed for all the observed objects to 45 min. The data reduction was performed by a pipeline designed specifically for this survey \citep[see details in][and \citealt{2014arXiv1409.8302G} for an update to the second data release]{2012A&A...538A...8S, 2013A&A...549A..87H}. 

To extract the flux from the ionized gas emission lines, we used the  FIT3D package~\footnote{http://www.caha.es/sanchez/FIT3D/} \citep{2006AN....327..167S, 2011MNRAS.410..313S}. We obtain these fluxes for the following lines: H$\alpha$, H$\beta$, [\ion{O}{iii}]~$\lambda\lambda$ 4959, 5007, and [\ion{N}{ii}]~$\lambda\lambda$5755, 6584. The analysis, including errors, was performed on a spaxel-by-spaxel basis, which allows us to create flux maps for each of these emission lines, once subtracted the underlying stellar population (see an example of emission-line maps in Fig.\,\ref{map}). We have derived the EW(H$\alpha$) in each spectrum by dividing the H$\alpha$ flux by the average of the continuum emission adjacent to  H$\alpha$ emission line. This average continuum was derived by measuring the mean flux of the continuum in spectral windows of 30 \AA\, centered at blue-shifted and redshifted 100\AA\, away from the  H$\alpha$ emission line.

\subsection{EW(H$\alpha$) and oxygen abundance measurements}    
\label{sec:Method}
Figure\,\ref{map} shows an example of the wealth of the data provided by the data cubes. Even for a single galaxy, several structures can be observed in the emission line and the EW(H$\alpha$) maps, which is a proxy for the specific star formation rate (sSFR). However, in this study we limited ourselves to the integrated properties for each galaxy. Studies exploring the star formation and ionized gas two-dimensional distributions in interacting galaxies will be left for future works. Thus from the emission line flux maps, we obtained the integrated properties for each galaxy. We selected those spaxels with errors smaller than 10\% of their flux within two apertures centered on the optical nucleus: a central and an extended aperture.

To study the impact of the scale used to derive the properties of the ionized gas, we selected two different scales for the diameter of the central and extended apertures: a projected (5 and 30 arcsecs) and the effective radius (0.3 and 2.5 R$_{\mathrm{eff}}$). We derived the effective radius for each galaxy using the same methodology as outlined in the Appendix~A of \cite{2014A&A...563A..49S}. For the projected scale, we chose the size of the central aperture to be approximately twice the size of the spatial PSF in the survey (FWHM $\sim$ 2.5 arcsec). The central region in both samples covers between $\sim$ 0.6 and $\sim$ 3 kpc. The extended aperture was chosen to cover a wide portion of the galaxies. In terms of the Petrosian radius we covered a generous portion of the galaxies (i.e, 1.5 $\lesssim$ $r_P$ $\lesssim$ 8). The central aperture in units of R$_\mathrm{eff}$ is, in most of the cases, larger than the spatial PSF. It is also comparable to the central projected aperture. In most of the galaxies, the extended aperture in units of  R$_\mathrm{eff}$ covers the entire FoV (see Fig.\,\ref{map}). In a few cases where the two companions in a binary system are covered by a single FoV, this aperture includes both galaxies (e.g., NGC~169). Our goal has been to compare how the properties of the ionized gas in the central and extended regions of interacting galaxies change when we compare it with a control sample (see Secs.\,\ref{sec:BPT}, \ref{sec:EW}, and \ref{sec:Z}). Nevertheless, our data also allow us to study the change of the ionized gas metallicity on different galactic scales (see Sec.\ref{sec:Z_r}).

To obtain the EW(H$\alpha$) for each galaxy, we averaged the individual spaxel absolute values included within each of the apertures. In the case of the observed oxygen abundance in each aperture, we used the empirical calibration O3N2 derived in \cite{2013A&A...559A.114M}, which makes use of the logarithm of the fluxes ratios  $\log{([\ion{O}{iii}]~\lambda5007/\mathrm{H}\beta)}$  and $\log{([\ion{N}{ii}]~\lambda6584/\mathrm{H}\alpha)}$. The O3N2 empirical calibrator has been proven to be a robust estimator of the metallicity when it is no possible to have its direct estimation by auroral lines \citep{2010A&A...517A..85L}. In Sec.\,\ref{sec:Z_cal} we study the impact of this oxygen abundance calibrator by contrasting our measurements using the above calibrator with the one presented by \cite{2004MNRAS.348L..59P}.

% ==========================================================================================
\section{Results}
\label{sec:results}
\subsection{BPT diagram}
\label{sec:BPT}
\begin{figure}[!htb]
 \includegraphics[width=\linewidth]{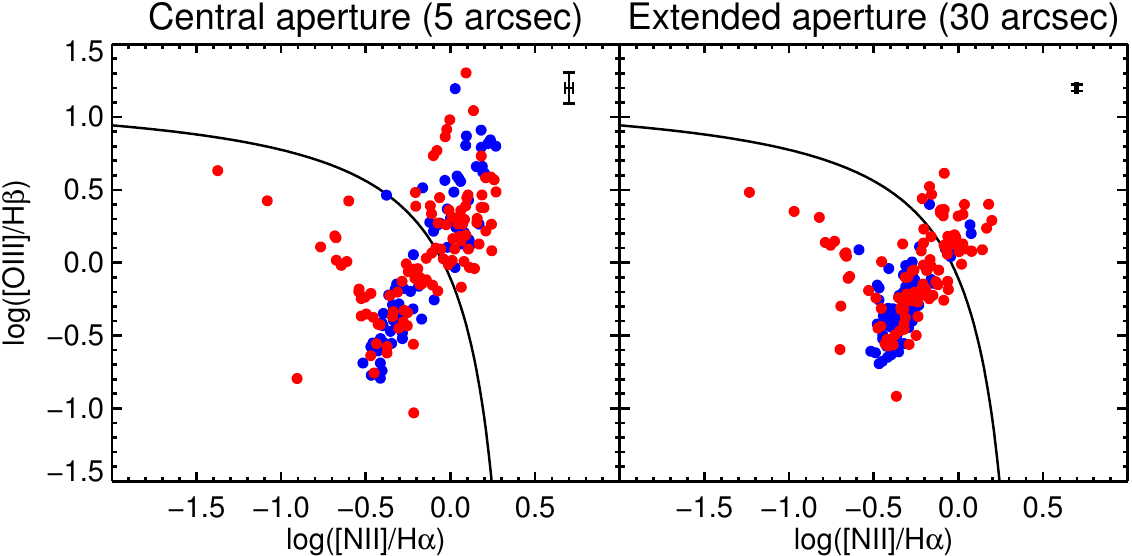}      
 \includegraphics[width=\linewidth]{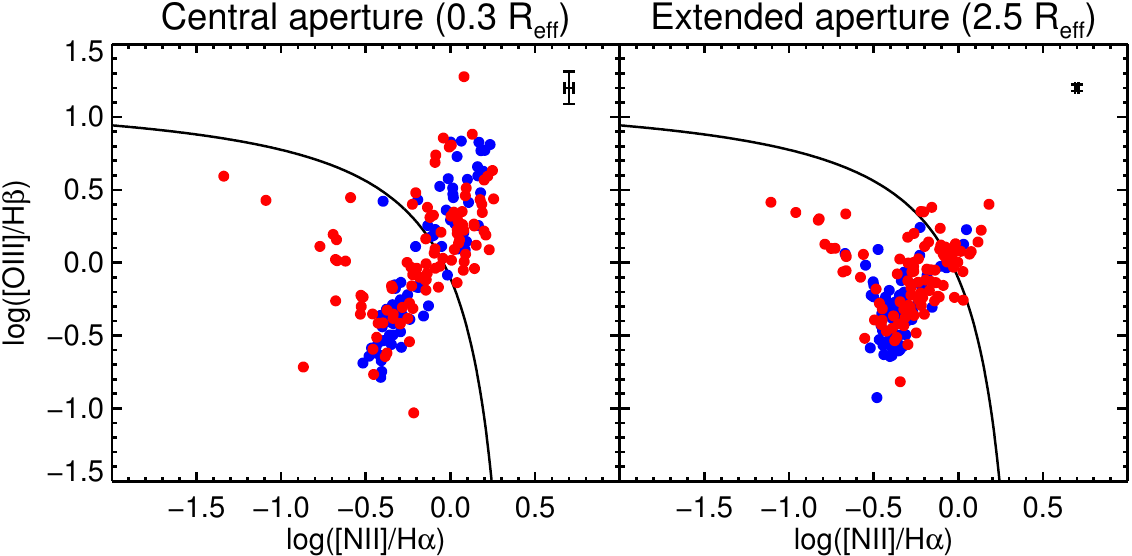}        
\caption{\label{BPT}BPT diagnostic diagrams for the control (blue points) and interacting (red points) samples measured within a central (left panels) and extended (right panels) apertures centered in the optical nucleus. Fluxes in top panels are derived using apertures on an arcsec scale, while fluxes in the bottom panel are obtained using an effective radius scale. The solid lines in each panel represent the division between star-forming and non-star-forming galaxies presented in \cite{2001ApJ...556..121K}. The typical uncertainty is plotted at the top of each panel.}
\end{figure}

We plot in Fig.~\ref{BPT} for each aperture and scale described in Sec.\,\ref{sec:Method} the classical diagnostic diagram using the line ratios $\log{([\ion{O}{iii}]~\lambda5007/\mathrm{H}\beta)}$ vs $\log{([\ion{N}{ii}]~\lambda 5755/\mathrm{H}\alpha)}$ \citep[][ BPT diagram hereafter]{1981PASP...93....5B}  in order to determine the fraction of star-forming galaxies in each of the samples for a given aperture size. We also included the demarcation line described by \cite{2001ApJ...556..121K}. Star-forming objects lie below this line, while AGN-powered sources are mostly located above it. On both size scales, the trends in each aperture are similar. For the central apertures (see left panels of Fig.\,\ref{BPT}) in both samples, a large number of galaxies are spread over the righthand side of the BPT diagram, from the star-forming to the AGN zone. However, some interacting objects lie in the so-called left branch (i.e., $\log{(\mathrm{[\ion{N}{ii}]}\lambda6583)/\mathrm{H}\alpha})$ $\lesssim$ - 0.6), associated with star-forming galaxies. In the extended apertures (right panels of Fig.\,\ref{BPT}), a larger percentage of galaxies lie in the star-forming zone, indicating that the integrated flux includes a larger number of \ion{H}{ii} regions than in the central aperture. 

We note that for both aperture sizes in these BPT diagrams, there is a non-negligible number of interacting galaxies located in the left  branch, whereas none of the control sample galaxies are located in this zone. As explained in \cite{2014A&A...563A..49S}, objects located in this branch own a high percentage of young stellar population. For the central aperture, these galaxies are included in pairs or interacting systems (10 objects on both aperture scales). This also holds for the extended aperture (13 and 12 objects for the 30 arcsec and 2.5 effective radius apertures, respectively). Since these galaxies are expected to have a significant percentage of young stellar population, these results suggest an increment in the SFR in interacting galaxies with respect to the control sample, particularly in binary systems. In the following section, we study the impact on the change in the sSFR by the interaction/merger across the galaxies by means of the spatially resolved EW(H$\alpha$).
\subsection{Central and extended EW(H$\alpha$)}
\label{sec:EW}

\begin{figure}[!htb] 
\includegraphics[width=\linewidth]{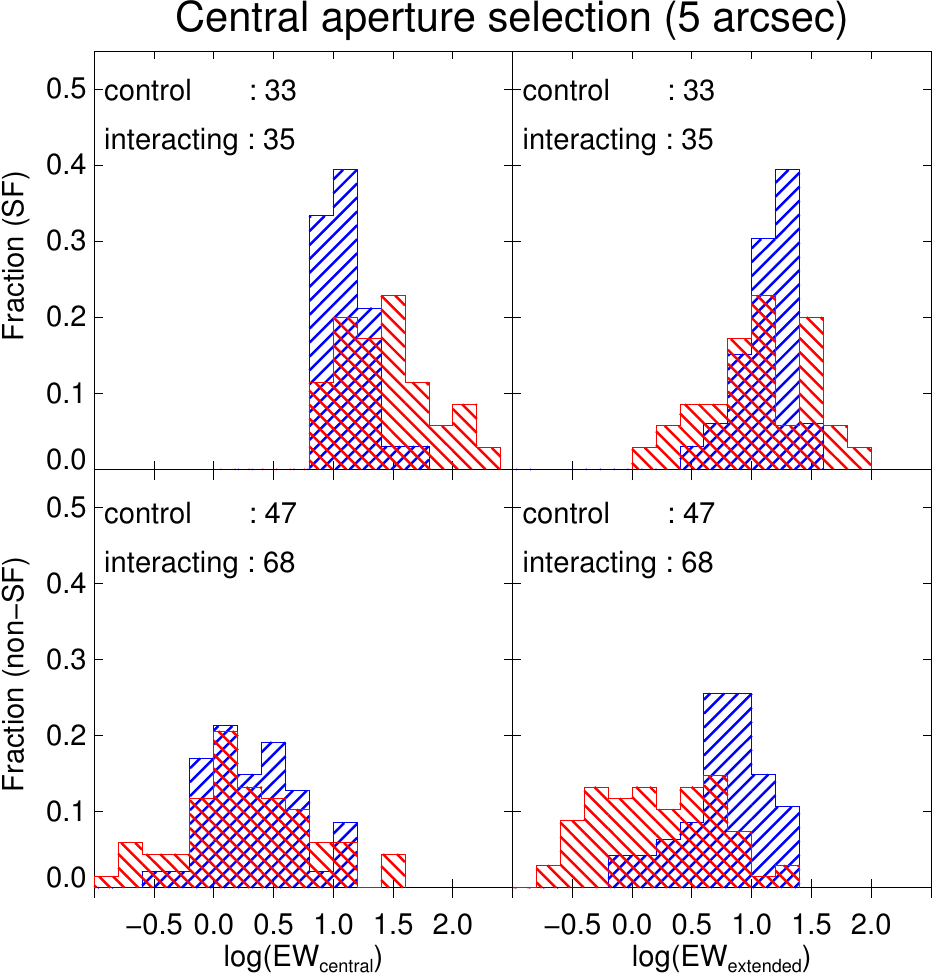}
\caption{\label{EW_arc_cen} Distributions of the integrated EW(H$\alpha$) for the interacting (red) and control (blue) samples using an aperture size of 5 arcsec. Top and bottom panels show the distributions of the star and non-star-forming galaxies, respectively. Left and right panels show the EW(H$\alpha$) distributions measured using the 5 (EW$_{\mathrm{central}}$) and 30 (EW$_{\mathrm{extended}}$) arcsec apertures, respectively. For this classification we used the integrated emission line fluxes ratios presented in Sec.\,\ref{BPT} using the same selecting aperture size.}
\end{figure}

Once we distinguished between star and non-star forming galaxies for each aperture size in both the control and interacting samples, we explore how the EW(H$\alpha$) changes on the different scales of these galaxies. The subsample of star-forming galaxies is selected in both samples following the criteria used by \cite{2014A&A...563A..49S}: objects with line ratios below the Kewley demarcation line in the BPT diagnostic diagram in  Fig.\,\ref{BPT} and EW(H$\alpha$) larger than 6 \AA. Our spatially resolved data allowed us to study the distribution of the EW(H$\alpha$) at different aperture sizes in galaxies classified as star-forming and non-star-forming galaxies depending on the ionized gas fluxes in each aperture. From Fig.\,\ref{BPT} we note that  non-star-forming sample includes mostly AGN-like objects, suggesting that the fraction of objects with a LINER-like spectrum is negligible. 

\begin{figure}[!htb] 
\includegraphics[width=\linewidth]{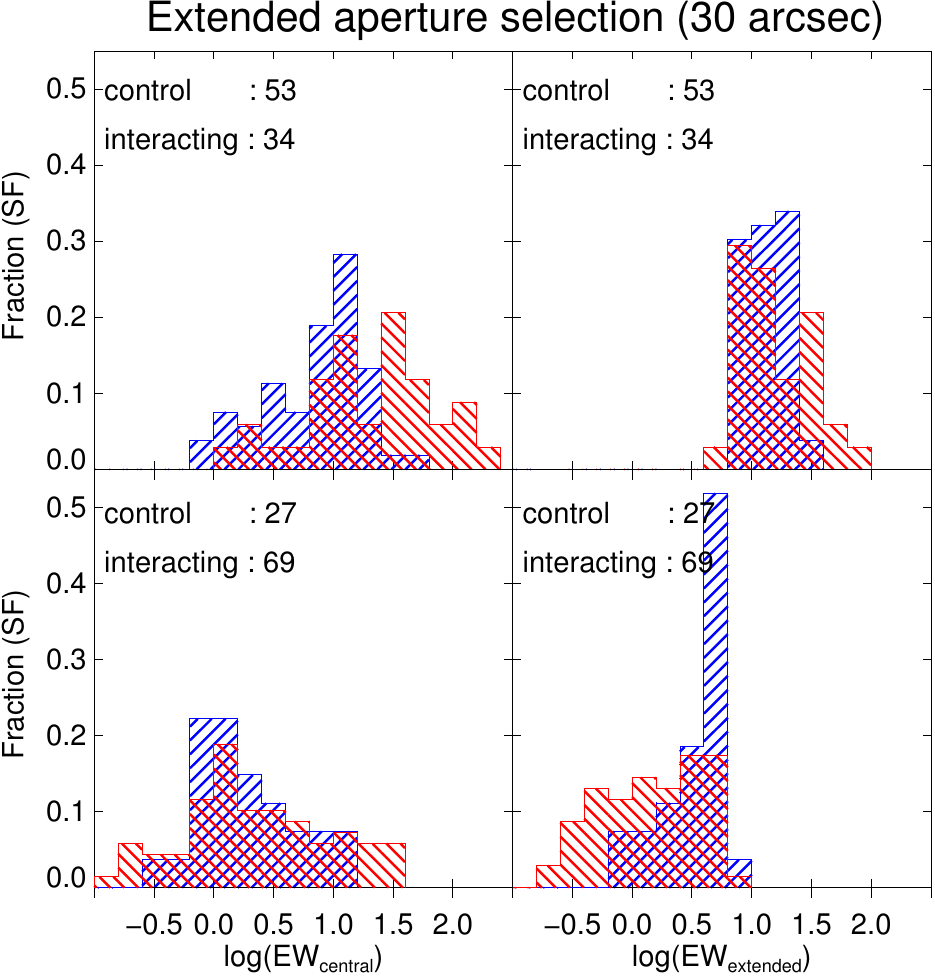}
\caption{\label{EW_arc_ext} Distributions of the integrated EW(H$\alpha$) using an aperture size of 30 arcsec. As in Fig.\,\ref{EW_arc_cen} blue and red histograms represent control and interacting EW(H$\alpha$) distributions, respectively. Top and bottom panels show the distributions for star-forming and non-star-forming galaxies, respectively.
}
\end{figure}

\begin{figure}[!htb] 
\includegraphics[width=\linewidth]{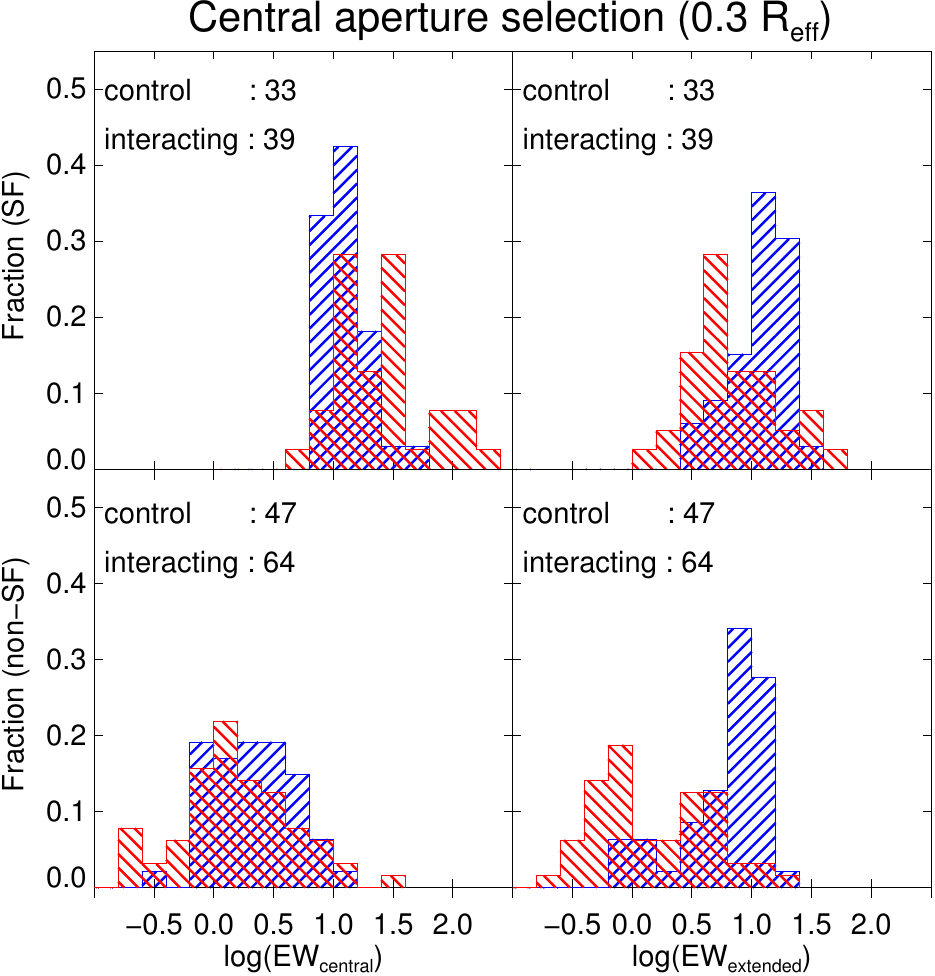}
\caption{\label{EW_reff_cen} Distributions of integrated EW(H$\alpha$) for the interacting (red) and control (blue) galaxies. The distribution of the panels is similar to Fig.\,\ref{EW_arc_cen}. For these histograms we use an aperture size of 0.3 R$_{\mathrm{eff}}$.}
\end{figure}

\begin{figure}[!htb] 
\includegraphics[width=\linewidth]{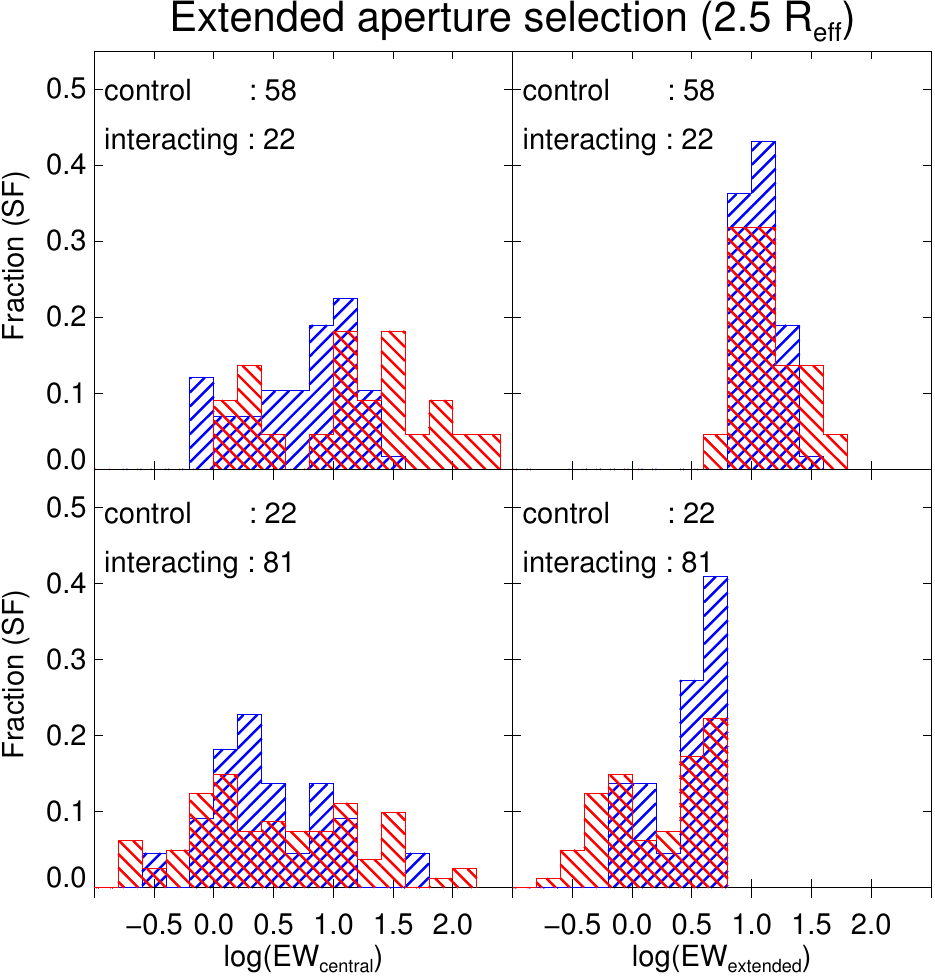}
\caption{\label{EW_reff_ext} Distributions of integrated EW(H$\alpha$) for the interacting (red) and control (blue) galaxies using an aperture size of 2.5 R$_{\mathrm{eff}}$. The distribution of the panels is similar to Fig.\,\ref{EW_reff_cen}. }
\end{figure}

In Fig.\,\ref{EW_arc_cen} we plot the $\log$(EW(H$\alpha$)) distributions of star-forming and non-star-forming galaxies for both samples at the 5 and 30 arcsec apertures (EW$_{\mathrm{central}}$ and EW$_{\mathrm{extended}}$, respectively) for galaxies classified as star-forming and non-star-forming galaxies in the 5 arcsec aperture (see Sec.\,\ref{BPT}). In Fig.\,\ref{EW_arc_ext} we plot the same distributions using the integrated fluxes within the 30 arcsec aperture as classification. We repeat the same exercise using the 0.3 and 2.5 R$_{\mathrm{eff}}$ aperture sizes. (see Figs.\,\ref{EW_reff_cen} and \ref{EW_reff_ext}). We note that the EW(H$\alpha$) distributions in both samples using either the arcsec or effective radii scales are similar for the central and extended aperture sizes, respectively. To make a reasonable comparison with previous results using single-fiber spectroscopic, we focus our analysis in the EW(H$\alpha$) derived from the arcsec apertures. 

Although the fraction of galaxies selected in the central aperture as star forming is similar for the control and interacting samples (33 and 35 objects, respectively), when we compare their  EW$_{\mathrm{central}}$ distributions, we find that a significant fraction of interacting galaxies (10/35) present higher values than does the entire control sample (see top left panel of Fig.\,\ref{EW_arc_cen}). Most of these objects are either in close pairs or in binary systems with evident signatures of interaction (33/35). A majority of these interacting galaxies are late types (28/35). We also find clear statistical differences between these two samples. The median $\log$(EW$_{\mathrm{central}}$) of the star-forming interacting subsample is larger than the median of the control sample ([1.41 $\pm$ 0.07] $\log$(\AA)\, and [1.09 $\pm$ 0.03] $\log$(\AA), respectively) \footnote{Errors in the medians and standard deviations are obtained from a bootstrapping method.}. These interacting galaxies cover a wide range of logEW(H$\alpha$) in comparison to the control sample (standard deviations of [0.38 $\pm$ 0.07] $\log$(\AA)\, and [0.39 $\pm$ 0.07] $\log$(\AA), respectively). Moreover, a Kolmogorov-Smirnoff test (KS-test, hereafter) reveals that these two samples are not likely to be drawn from the same parent sample ($p_{KS}$ = 0.001). We transform the EW(H$\alpha$) to sSFR by means of the empirical relation presented by \cite{2013A&A...554A..58S}. For the interacting star-forming galaxies, the median sSFR is (2.3 $\pm$ 0.2) $\times$ 10$^{-10}$ yr$^{-1}$, while the median sSFR for the control subsample is approximately half the interacting value ([9.3 $\pm$ 0.9] $\times$ 10$^{-11}$ yr$^{-1}$). 

The distributions of EW$_{\mathrm{extended}}$ for galaxies selected as star forming in the central aperture are fairly similar (see top right panel of Fig.\,\ref{EW_arc_cen}), and median values are similar in both samples ([1.07 $\pm$ 0.01] $\log$(\AA)\, and [1.18 $\pm$ 0.02] $\log$(\AA), respectively). The moderate sSFR enhancement we observe in central regions of interacting galaxies ($\sim$ 2.5 times) has been reported by previous statistical studies \citep[e.g.,][]{2006AJ....131.2004K, 2008AJ....135.1877E, 2009ApJ...698.1437K, 2013MNRAS.433L..59P}. However, we point out that those studies covered a portion that was larger than the central region. In fact, their aperture sizes are similar to our extended aperture where we find a moderate reduction of the sSFR compared to isolated galaxies ($\sim$ 0.74).
 % This confirms previous statistically studies where the central star formation activity is enhanced moderately by interaction .

In comparison to the star-forming galaxies, the non-star-forming galaxies selected by the central aperture in both samples present a rather similar distribution toward low EW$_{\mathrm{central}}$ values (see bottom left panel of Fig.\,\ref{EW_arc_cen}). We would like to point out here that these subsamples include objects located in the AGN zone in the BPT diagram or those with EW(H$\alpha$) smaller than 6\AA. Although it is beyond the scope of this study, the similarity between these distributions deserve more attention so we can understand the impact of the interaction in triggering the nuclear activity of galaxies. We find however, a clear difference between the distributions of EW$_{\mathrm{extended}}$ for the same sample of galaxies (see bottom right panel of Fig.\,\ref{EW_arc_cen}). The median EW$_{\mathrm{extended}}$ of the interacting sample is four times smaller than the control one ([0.20 $\pm$ 0.08] $\log$(\AA)\, and [0.82 $\pm$ 0.05] $\log$(\AA), respectively).

The control star-forming galaxies selected using the extended aperture almost doubles the interacting ones (see top left panel of Fig.\,\ref{EW_arc_ext}). The median $\log$(EW$_{\mathrm{central}}$) for the interacting subsample is almost three times larger than the median of the control sample ([1.41 $\pm$ 0.1]$\log$(\AA)\, and [0.99 $\pm$ 0.03] $\log$(\AA), respectively). For the same subsample of star-forming galaxies, EW$_{\mathrm{extended}}$ is similar for the interacting and control samples ([1.11 $\pm$ 0.02] $\log$(\AA), see top right panel in Fig.\,\ref{EW_arc_ext}). A KS-test reveals that these two samples could be drawn from the same parent sample ($p_{KS}$ = 0.3). This suggests that the net enhancement of the sSFR occurs mainly in the central region of the interacting galaxies rather than in the outer regions.

In the extended aperture, the selection of non-star-forming galaxies yields almost three times more interacting objects than control ones. The distribution, as well as the median of EW$_{\mathrm{central}}$, is similar for both samples ([0.18 $\pm$ 0.07] $\log$(\AA), see bottom left panel in Fig.\,\ref{EW_arc_ext}). On the other hand, EW$_{\mathrm{extended}}$ present significant differences between the interacting and control samples (see bottom right panel in Fig.\,\ref{EW_arc_ext}). The median EW$_{\mathrm{extended}}$ for this interacting subsample is three times smaller than the control one ([0.17 $\pm$ 0.06] $\log$(\AA) and [0.64 $\pm$ 0.03] $\log$(\AA), respectively). In fact, a KS-test indicates that these two distributions are not likely to be drawn from the same parent sample ($p_{KS}$ = 0.001).

Our analysis shows that regardless of the criteria we use to select star-forming galaxies, the interacting sample has a consistent moderate enhancement in the central sSFR when compared with isolated star-forming objects. On extended scales, both samples present similar distributions of EW$_{\mathrm{extended}}$ and, depending on the aperture used to classify the star-forming galaxies, similar or moderately suppressed total sSFR. Most of these interacting objects are late-type galaxies in pairs or merging systems. These results suggest that  moderate enhancement ($\sim$ 2-3 times larger) in the sSFR only occurs in the central region of interacting galaxies. This scenario is consistent with gas been funneled to the central region of galaxies owing to the interaction: large amounts of gas (subsequently enhanced sSFR) is found in the central region of interacting galaxies, whereas the integrated sSFR remains similar or moderately suppressed. Our spatially resolved study agrees with the numerical simulations presented by \cite{2015MNRAS.448.1107M}. They find that the increment of star formation in merging galaxies is observed within the central kpc, while in outer galacto-centric radii, the activity is moderately suppressed. Our results also agree with studies of supernovae radial distributions \citep[e.g., ][]{2012MNRAS.424.2841H, 2012A&A...540L...5H} that find Type-Ibc supernovae, those more correlated to the ongoing star formation \citep[][]{2014A&A...572A..38G}, are more centrally concentrated in disturbed/interacting galaxies.

In galaxies considered as non-star-forming, the central EW(H$\alpha$) distribution is similar between interacting and isolated objects, whereas the extended EW(H$\alpha$) is systematically smaller for the interacting sample. This points out that the process responsible for the observed central EW(H$\alpha$) in non-star-forming galaxies for both samples could be similar. From the BPT diagrams (see Fig.\,\ref{BPT}), this process could involve only AGN activity. On the other hand, low values of the extended EW(H$\alpha$) for interacting galaxies could imply either a low budget of gas available to be ionized or a weak radiation field across the merging galaxies. Since these galaxies are the complement to the star-forming subsample, they cover a wide range of interaction stages and morphological types. A further division of the non-star-forming galaxies EW(H$\alpha$) in the above properties led a few objects in each bin, making it difficult to derive reliable results. We require a large sample of merging galaxies in particular in the post-merger and remnant stages to derive statistical meaningful conclusions about the mechanisms responsible for the observed trends in the central and extended EW(H$\alpha$) in non-star-forming merging galaxies.   
 
\subsection{Central oxygen abundances} 
\label{sec:Z}
\begin{figure}
\minipage{0.5\linewidth}         
\includegraphics[width=\linewidth]{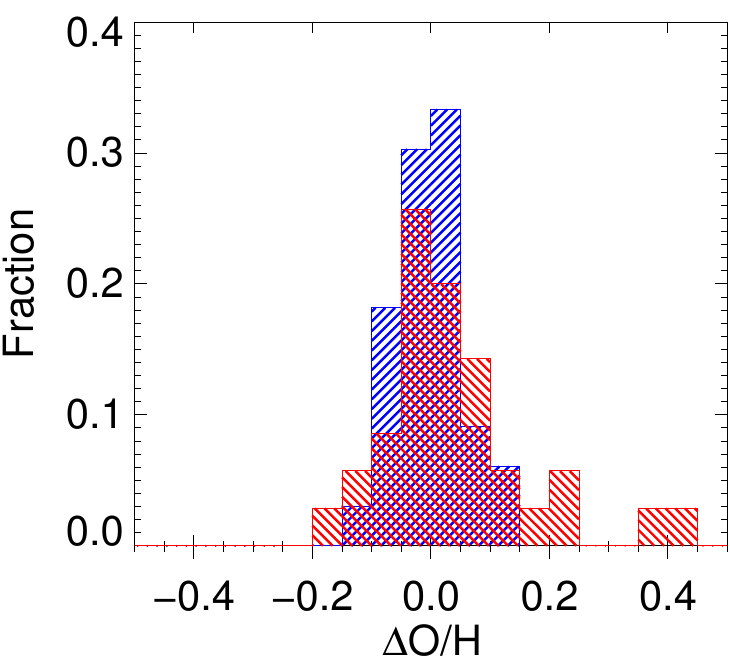}
\endminipage
\minipage{0.5\linewidth}         
\includegraphics[width=\linewidth]{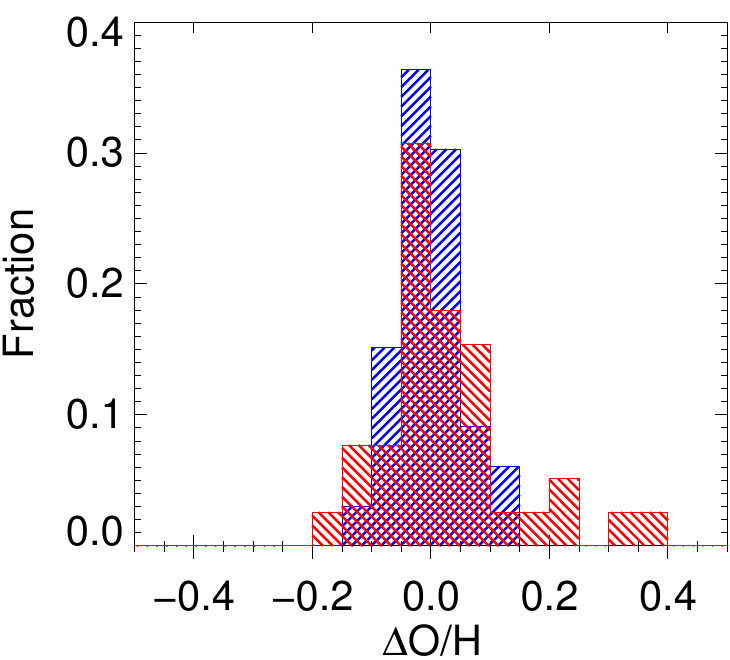}
\endminipage
\caption{\label{Z} \emph{Left:} Distributions of $\Delta$O/H in the central (5 arcsec) aperture for the interacting (red) and the control (blue) star-forming galaxies, see text for details. \emph{Right:} Distributions of $\Delta$O/H using a central aperture of 0.3 R$_{\mathrm{eff}}$.}
\end{figure}
Observational studies suggest that the enhancement in the star formation activity in the central region of interacting galaxies is connected with a dilution of the oxygen abundance \citep[e.g.,][]{2006AJ....131.2004K, 2008AJ....135.1877E, 2014A&ARv..22...71S, 2014ApJ...783...45S}, which in turns contributes to the scatter of the M-Z relation. This decrement is associated with the supply of metal-poor gas to the central region via inflows. To test this scenario, we studied the scatter between the observed metallicity and the M-Z relation for a given stellar mass in star-forming galaxies. From the emission
line fluxes, we derived the oxygen abundance of each galaxy classified in Sec.\,\ref{sec:EW} as star-forming in the central aperture for both the interacting and control samples. Then, we provided the difference (or scatter) between the abundance determined from the flux ratios and the one expected from the M-Z relation ($\Delta$O/H) derived by \cite{2013A&A...554A..58S}. For the objects included in the CALIFA survey, we used the stellar masses presented by \cite{2014A&A...569A...1W}. For the companions not included in the survey, stellar masses were derived in a similar fashion to the CALIFA galaxies. As explained in  \cite[][]{2013A&A...554A..58S}, this empirical M-Z relation is derived for metallicities at one effective radius. Nevertheless, we can transform this abundance at different radii, in particular to the central aperture radius, by means of the metallicity gradient in disk galaxies \citep{2014A&A...563A..49S}. 

In  Fig.~\ref{Z} we plot the distributions of $\Delta$O/H for both the interacting and control samples in the central aperture (5 arcsec and 0.3 R$_{\mathrm{eff}}$, left and right histograms, respectively). The distribution for the control sample in the 5 arcsec aperture is centered well on zero, with a median value of (0.001\,$\pm$\,0.006) dex and a rather narrow distribution ($\sigma\Delta\mathrm{O/H}_{\mathrm{control}}$ = 0.056\,$\pm$\,0.006 dex). On the other hand, the median of the distribution from the interacting sample is slightly larger than the one from the control sample (0.03\,$\pm$\,0.01 dex). However, the interacting sample has a wider range of values in comparison with the control sample ($\sigma\Delta \mathrm{O/H}_{\mathrm{merger}}$ = 0.10\,$\pm$\,0.01 dex). A KS-test indicates that these two samples are likely to be drawn from the same parent sample distribution ($p_{KS}$ = 0.32). We find a similar distribution for the 0.3 R$_{\mathrm{eff}}$ aperture with median values of (-0.001\,$\pm$\,0.006) dex and (0.021\,$\pm$\,0.02) dex for the control and interacting samples, respectively. The interacting sample is slightly more metal rich than the control sample (0.02\,$\pm$\,0.02 dex). This indicates that in central regions, star-forming merging galaxies have similar oxygen abundance to the control sample. Even though both apertures provide similar results, the most reliable one comes from the 0.3 R$_{\mathrm{eff}}$ aperture size since it covers a similar physical region in all the galaxies in both samples. This first IFS census suggests that even though there is an enhancement of the sSFR in the central region of interacting galaxies, the oxygen abundance remains similar to the one found in isolated galaxies. In other words, the scatter observed in the M-Z relation does not seem to be linked to the star formation rate of the galaxies presented in this study.

Our results differ from the single-fiber spectroscopic studies carried out with the SDSS galaxy-pair sample \citep{2008AJ....135.1877E}. They found a lower oxygen abundance of $\sim$ 0.05 - 0.1 dex between their galaxy-pair samples and control sample. One evident reason for the disparity between these two studies could be attributed to the number of objects in each sample. SDSS pairs and control samples included 1716 galaxies in binary systems and 40095 galaxies, respectively, whereas our samples of star-forming galaxies include 33 and 35 objects in both interacting and control samples, respectively. Despite this difference in the size of the samples, we are able to account for the enhancement in the central SFR of interacting galaxies (see Figs.\,\ref{EW_arc_cen} and \ref{EW_reff_cen}), in agreement with the SDSS studies, but not for the dilution in central metallicities. We note that our reported small variations in the metallicity in the central aperture may be biased by aperture effects or even the calibration used to derived the metallicity. In the following two sections, we explore the possible bias in our results for these two factors. 
\begin{figure*}
\minipage{0.5\linewidth}         
\includegraphics[width=\linewidth]{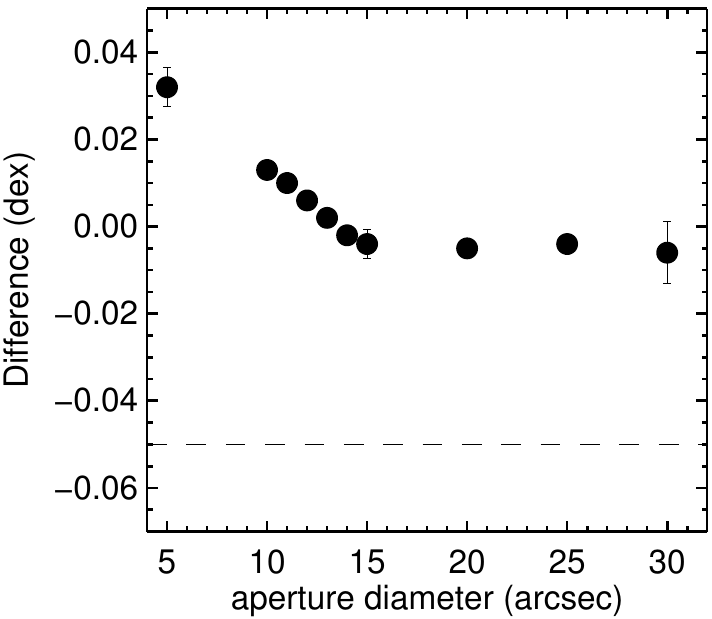}
\endminipage
\minipage{0.5\linewidth}         
\includegraphics[width=\linewidth]{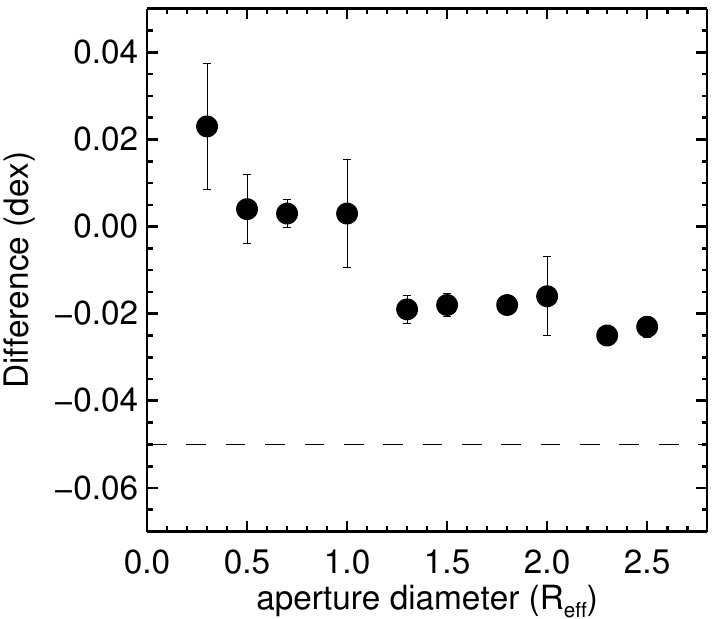}
\endminipage
\caption{\label{Z_r} Differences in the median values of $\Delta$O/H between the interacting and control samples at different aperture sizes. \emph{Left:} The aperture sizes are measured in arcsecs. \emph{Right:} Aperture scales are measured in effective radius. The dashed line represents the difference reported by \cite{2008AJ....135.1877E}. Errors bars are determined using bootstrapping in both samples. On any of the scales, the differences reach the values found in single-fiber spectrocopic studies.} 
\end{figure*}
\subsection{Metallicity as function of the aperture size} 
\label{sec:Z_r}
Our IFS data allows exploring the effect of measuring the gas metallicity for different aperture sizes and scales. In particular, we can study whether the lower abundances presented in single-fiber studies for the star-forming interacting galaxies \citep[e.g.,][]{2008AJ....135.1877E} could be caused by aperture effects.  More important, we will have an estimation of how the metallicity changes at different galactic scales. In Fig.\,\ref{Z_r} we plot the differences in the median metallicity value for the interacting and control star-forming galaxies at different aperture sizes. We plot these differences for two aperture scales in arcsecs and in effective radius. The sample of galaxies selected for these plots are those classified as star-forming galaxies in the central aperture of each scale.

For the aperture scale in arcsecs, we observe that the difference in metallicities between the two samples decreases as the aperture size increases, reaching a plateau at $\sim$ 15 arcsec. After this size the differences in metallicity are rather constant for different aperture sizes ($\sim$ -0.01 dex). Using the effective radius of each galaxy as aperture scale, the difference in metallicity for the two samples decreases as the size of the aperture increases. For the largest aperture (2.5 R$_{\mathrm{eff}}$), the interacting galaxies shows a lower metallicity of $\sim$ 0.02 dex with respect to control galaxies. Although there is a decrement in the metallicity of merging galaxies for large apertures, we do not observe the difference presented by previous spectroscopic studies \citep[e.g.,][]{2006AJ....131.2004K, 2008AJ....135.1877E}. Even more in the physical motivated aperture (R$_{\mathrm{eff}}$), we find rather similar metallicities in the central regions with hints of a dilution of metallicities of the interacting galaxies on larger scales. This suggests that despite the inflows of ionized gas induced by the interaction, the metallicity in the central regions of interacting galaxies presents similar properties to those for isolated galaxies. On large scales, there seems to be a dilution in the metallicity of interacting objects.

\subsection{Metallicity as function of the abundance calibrator} 
\label{sec:Z_cal}

\begin{figure}[!!!htb]
\includegraphics[width=\linewidth]{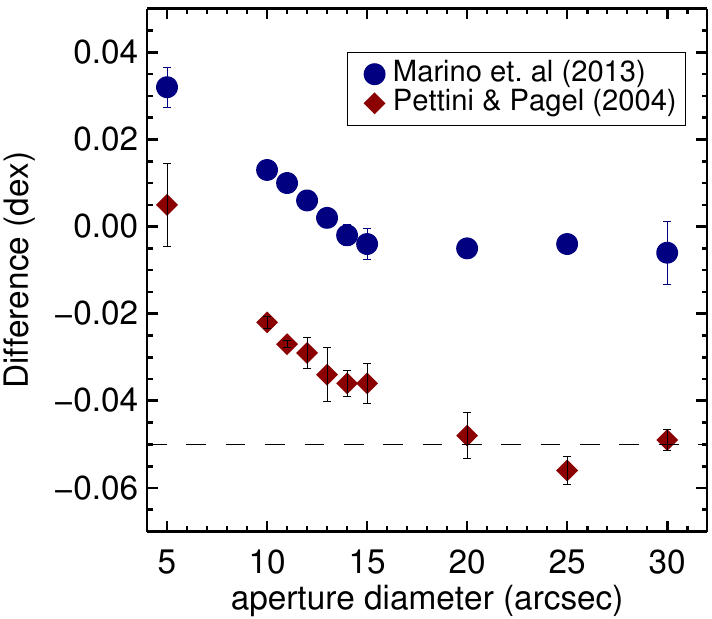}
\caption{\label{Z_cal}Similar to Fig.\,\ref{Z_r}. Differences in the median values of $\Delta$O/H between the interacting and control samples at different aperture sizes. In this case we compare using apertures in arcsec units with the differences in metallicity using different calibrators. Blue circles represent the differences obtained from the O3N2 revisited calibrator used in this study given in \cite{2013A&A...559A.114M}. Red diamonds show the difference obtained from the calibrator presented by \cite{2004MNRAS.348L..59P}.  The dashed line represent the difference reported by \cite{2008AJ....135.1877E}. } 
\end{figure}

In general, the variations in metallicity between the control and interacting galaxies are not larger than 0.1 dex. This regime of small variations can be biased by several factors. In this section we explore and quantify the impact of these differences by the abundance calibrator. In Fig.\,\ref{Z_cal} we plot the differences as a function of aperture sizes using the improved empirical calibration for the O3N2 indicator given by \cite{2013A&A...559A.114M}. We overplot these differences using a different calibration provided by \cite{2004MNRAS.348L..59P}. As we note above, the differences between the interacting and control sample are very small when using the improved calibrator, flattening out at $\sim$ 15 arcsecs  to $\sim$ -0.01 dex. However, when we use the calibrator from \cite{2004MNRAS.348L..59P}, we find that in larger apertures, the interacting galaxies have lower metallicities than control galaxies. Even more at $\sim$ 20 arcsec the differences in metallicity reach the value found by \cite{2008AJ....135.1877E}.

To explain this apparent difference between our results and those using another calibrator, we note that our sample is closer than the sample presented by \cite{2008AJ....135.1877E}. At the median redshift of the SDSS-pair sample ($z \sim$ 0.06, see their Fig.2), an aperture of 3 arcsec is equivalent to central regions of $\sim$ 3.5 kpc. On the other hand, an aperture of the 5 arcsec used to define the central region in this study covers a central portion of $\sim$ 1 kpc at the median redshift of our sample ($z \sim$ 0.01). As a consequence, the SDSS-pair single-fiber observations cover a wider region than the area covered by our central aperture. In fact, at the  20 arcsec aperture where the differences in metallicity (using the calibrator from \cite{2004MNRAS.348L..59P}) are similar to those found by \cite{2008AJ....135.1877E}, this aperture covers a region of $\sim$ 3.5 kpc at the median redshift of our sample. Although we did not use the same indicators for metallicity as for the single-fiber spectroscopic survey studies, we suggest that by using abundance calibrators presented in the literature, it is possible to obtain the same lower metallicity in interacting galaxies.

Finally we want to highlight that regardless of the indicator used to derive the oxygen abundance, the metallicity in the central regions for interacting galaxies is similar to the one derived for the control sample.

\section{Conclusions}
\label{sec:conclusions}
We used the spatially resolved information provided by the CALIFA survey to carry out the first statistical study of the impact of the merger event on the sSFR and the oxygen abundance on different galactic scales for a sample over 100 galaxies at different stages of interaction. Moreover, this survey allows us to homogeneously determine the same properties in a control sample of 80 non-interacting galaxies.
We separated the galaxies in both samples between star-forming and non-star-forming galaxies (see Secs.\,\ref{sec:BPT} and \ref{sec:EW}). For the first subsample, we find a (moderate) enhancement in the sSFR in the central region of interacting galaxies. However, in outer regions, the sSFR is similar to or moderately suppressed in comparison to the control sample (see Sec.\,\ref{sec:EW}). This agrees with previous observational studies \cite[e.g., ][]{2008AJ....135.1877E, 2012A&A...548A.117Y}, as well as with numerical simulations \cite[e.g.,][]{2013MNRAS.430.1901H, 2015MNRAS.448.1107M}. These studies indicate that tidally induced star formation becomes evident in the central region of the galaxies as a consequence of gas inflows. In this scenario, the new supply of metal-poor gas to the central region produces a dilute metallicity. In contrast to this picture, we find similar metallicities in the central region of star-forming, interacting galaxies to those derived for isolated galaxies (see Sec.\,\ref{sec:Z}). When we consider larger apertures (in effective radius units), we find hints of dilute metallicities in the interacting galaxies (see Sec.\,\ref{sec:Z_r}). Our results support the notion of a tight interplay between different physical processes in the central part of interacting galaxies. Although metal-poor gas inflows can be considered as the main process that affects the chemical evolution toward the center of interacting galaxies, there are some other processes to be taken into account that could enrich the ionized gas, such as stellar (or AGN) feedback or returned material into the ISM. The results presented here encourage IFS studies in larger samples of interacting galaxies in order to understand the evolution of the galactic chemical content and to quantify how different merging configurations can affect the merger-induced metallicity dilution.  
\section*{Acknowledgments}
The authors thank the referee for  the useful comments and suggestions. This study made use of the data provided by the Calar Alto Legacy Field Area (CALIFA) survey (http://www.califa.caha.es). It was based on observations collected at the Centro Astron\'{o}mico Hispano Alem\'{a}n (CAHA) at Calar Alto, operated jointly by the Max-Planck-Institut f\"{u}r Astronomie and the Instituto de Astrof\'{\i}sica de Andalucia (CSIC). CALIFA is the first legacy survey performed at the Calar Alto. The CALIFA collaboration would like to thank the IAA-CSIC and MPIA-MPG as major partners of the observatory, and CAHA itself, for the unique access to the telescope time and support  in manpower and infrastructures. The CALIFA collaboration also thanks the CAHA staff for their dedication to this project.
J.B-B. and B.G-L acknowledge support from the Plan Nacional de I+D+i (PNAYA) funding programs (AYA-2012-39408-C02-1 and AYA-2013-41656-P) of the Spanish Ministry of Economy and Competitiveness (MINECO). 
J.F.B.  acknowledges support from the Plan Nacional de I+D+i (PNAYA) funding programs from MINECO (AYA2013-48226-03-1-P, RAVET).
J.I.P. acknowledges financial support from the Spanish MINECO through grant AYA2010-21887-C04-01 and from the Junta de
Andaluc\'{\i}a Excellence Project PEX2011-FQM7058.
R.A.M is funded by the Spanish program of International Campus of Excellence Moncloa (CEI).
Support for L.G. is provided by the Ministry of Economy, Development, and Tourism's Millennium Science Initiative through grant IC120009, awarded to The Millennium Institute of Astrophysics, MAS. L.G. acknowledges support by CONICYT through FONDECYT grant 3140566. 
C.J.W. acknowledges support through the Marie Curie Career Integration Grant 303912.

\bibliographystyle{aa}
\bibliography{SFR_Z_final}

\end{document}